\PassOptionsToPackage{numbers,sort&compress}{natbib}
\documentclass{article}
\usepackage[preprint]{neurips_2026}

\usepackage[utf8]{inputenc}
\usepackage[T1]{fontenc}
\usepackage{hyperref}
\usepackage{url}
\usepackage{booktabs}
\usepackage{amsfonts,amsmath,amssymb}
\usepackage{nicefrac}
\usepackage{microtype}
\usepackage[table]{xcolor}
\usepackage{graphicx}
\usepackage{float}
\usepackage{flafter}
\usepackage{pifont}
\usepackage{multirow}
\graphicspath{{figs/out/}}
\newcommand{\AlphaStar}{0.1}
\newcommand{\GlidGen}{0.816}
\newcommand{\GlidSwap}{0.910}

\newcommand{\GlidHard}{0.636}

\newcommand{\GlidAllSixteen}{0.805}

\newcommand{\BtwoAllSixteen}{0.798}
\newcommand{\BtwoGen}{0.731}

\newcommand{\GenGain}{+0.084}
\newcommand{\ReenDelta}{-0.005}
\newcommand{\EffortAllSixteen}{0.790}

\newcommand{\SbiWild}{0.870}

\newcommand{\AlphaOneGen}{0.825}
\newcommand{\AlphaOneReen}{0.610}
\newcommand{\BtwoSeedMean}{0.788}
\newcommand{\BtwoSeedStd}{0.009}
\newcommand{\BtwoNSeeds}{8}
\newcommand{\DoseZeroGain}{+0.100}
\newcommand{\DoseZeroStd}{0.011}
\newcommand{\DoseOneGain}{-0.003}

\newcommand{\DoseZeroBtwoGen}{0.720}
\newcommand{\DoseOneBtwoGen}{0.873}
\newcommand{\DoseTwentyBtwoGen}{0.988}
\newcommand{\DoseOneImages}{113}
\newcommand{\DinoTtwoiMid}{0.892}
\newcommand{\DinoTtwoiDeep}{0.581}
\newcommand{\DinoStyleDeep}{0.863}

\newcommand{\VarBtwoGen}{0.0425}
\newcommand{\VarLidGen}{0.0078}
\newcommand{\VarCtrlGen}{0.0096}
\newcommand{\VarRatioLid}{5.5}
\newcommand{\VarRatioCtrl}{4.4}
\newcommand{\VarGainLid}{+0.118}
\newcommand{\VarGainCtrl}{-0.146}
\newcommand{\LatLid}{4.1}
\newcommand{\LatBtwo}{3.6}
\newcommand{\LatDca}{18.7}
\newcommand{\LatRigid}{8.0}
\newcommand{\EceT}{1.75}

\newcommand{\EceValT}{0.051}

\newcommand{\EceCrossT}{0.126}
\newcommand{\DcaSwap}{0.787}
\newcommand{\DcaGen}{0.595}
\newcommand{\DcaCelebdf}{0.847}
\newcommand{\LidGen}{0.791}
\newcommand{\LidSwap}{0.551}
\newcommand{\RigidReen}{0.729}
\newcommand{\RigidTtwoiRaw}{0.043}
\newcommand{\RigidTtwoiBlur}{0.779}
\newcommand{\RigidTtwoiJpegfifty}{0.300}
\newcommand{\LidTtwoiClean}{0.822}
\newcommand{\LidTtwoiJpegfifty}{0.783}
\newcommand{\LareTtwoiRaw}{0.102}
\newcommand{\IndistAtZero}{0.9931}
\newcommand{\IndistAtStar}{0.9920}
\newcommand{\KneeSoft}{0.2}
\newcommand{\ParetoMax}{0.3}
\newcommand{\ReenMissBy}{0.0013}
\newcommand{\CropDelta}{0.03}
\newcommand{\PairStyleSbiLo}{+0.188}
\newcommand{\PairStyleSbiHi}{+0.244}
\newcommand{\PairTtwoiSbiLo}{-0.087}
\newcommand{\PairTtwoiSbiHi}{-0.049}
\newcommand{\PairStyleEffLo}{+0.086}
\newcommand{\PairStyleEffHi}{+0.120}
\newcommand{\PairTtwoiEffLo}{+0.063}
\newcommand{\PairTtwoiEffHi}{+0.095}
\newcommand{\WinSbi}{8}
\newcommand{\LoseSbi}{4}
\newcommand{\WinEff}{3}
\newcommand{\LoseEff}{0}
\newcommand{\NumCitations}{72}
\newcommand{\BtwoSwap}{0.915}
\newcommand{\BtwoHard}{0.647}
\newcommand{\PoolOursGen}{0.816}

\newcommand{\PoolOursAll}{0.805}
\newcommand{\PoolAllfourGen}{0.810}

\newcommand{\PoolAllfourAll}{0.793}
\newcommand{\PoolDeepGen}{0.762}

\newcommand{\PoolMidGen}{0.795}

\title{GLID: Gated Local Intrinsic Dimension Repairs the\\Blind Spots of Face-Forgery Detectors}

\author{%
  Guang Yang\\
  University of California, Los Angeles\\
  \texttt{guangyang19@ucla.edu}\\
  \And
  Fengchen Liu\\
  University of California, Berkeley\\
  \texttt{fengchenliu@berkeley.edu}\\
}

\begin{document}

\maketitle

\begin{abstract}
Fine-tuned foundation-model detectors dominate face-forgery benchmarks, yet they stay blind to generator families absent from training. We present \emph{GLID}, a detector that repairs this blind spot with geometry instead of data. GLID treats the patch tokens of a single image as a sample from a manifold and estimates their \emph{local intrinsic dimension} (LID) at several depths of a frozen vision transformer. This 12-dimensional, training-free signal enters a fine-tuned detector through a confidence gate whose strength is calibrated purely in-distribution. On a 16-axis cross-generator benchmark, GLID reaches \GlidAllSixteen{} mean AUC, first among retrained state-of-the-art baselines and never significantly behind the strongest of them on any axis. It lifts the generation axes by \GenGain{} AUC while moving reenactment by only \ReenDelta{}. Two empirical laws explain the design. First, forged faces bend the token manifold at family-specific depths: GAN artifacts peak at the last layer, diffusion artifacts peak mid-network, and the pattern survives four backbones, three dimension estimators, and non-face imagery. Second, fine-tuning absorbs auxiliary gains exactly where training data covers: injecting 1\% target-family images erases a \DoseZeroGain{} gain, so geometric signals matter precisely where data is unavailable. The deterministic signal also cuts the cross-seed spread of accuracy \VarRatioLid$\times$. Code, preregistered analysis gates, and per-image scores accompany the paper.
\end{abstract}

\begin{figure}[t]
  \centering
  \includegraphics[width=\linewidth]{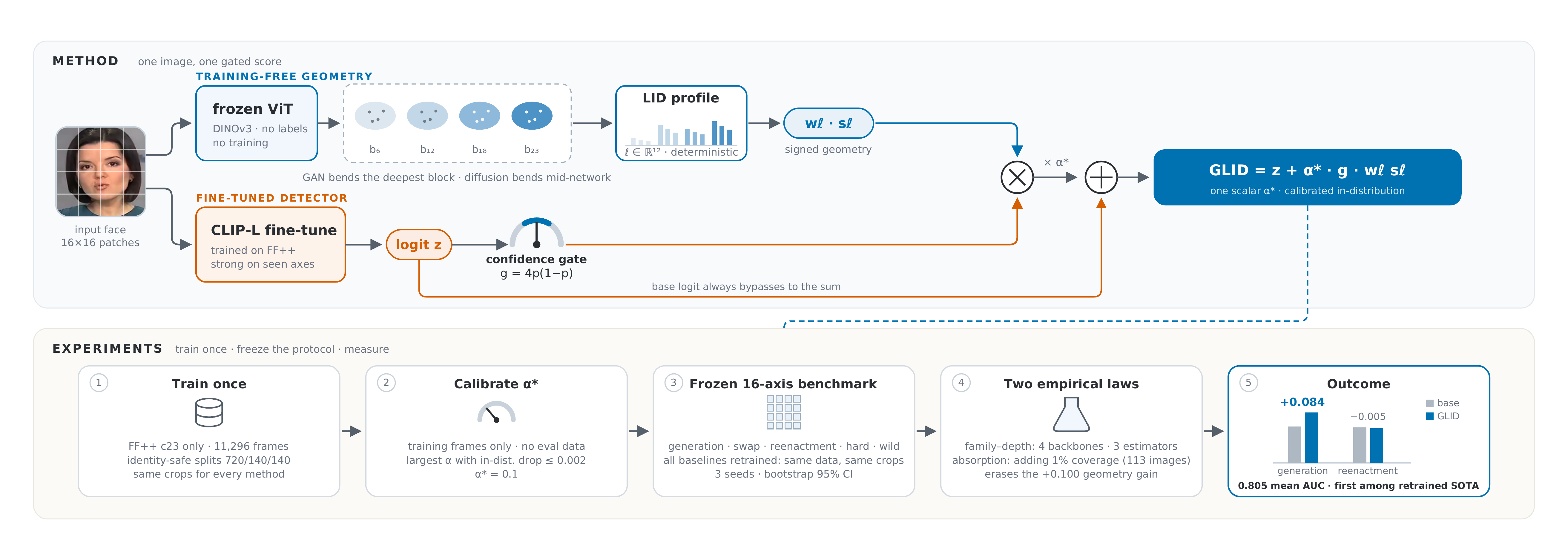}
  \caption{\textbf{GLID at a glance.} \emph{Method} (top): a frozen ViT yields a training-free, deterministic LID profile across four depths; a fine-tuned detector yields a logit and a confidence gate $g=4p(1-p)$; the gated geometric term joins the always-present base logit as $\mathrm{GLID}=z+\alpha^{*} g\, w_\ell s_\ell$. \emph{Experiments} (bottom): train once on FF++ c23, calibrate $\alpha^{*}$ purely in-distribution, evaluate on a frozen 16-axis benchmark with retrained baselines, and verify two empirical laws. Unseen generation axes gain \GenGain{} AUC while reenactment moves by \ReenDelta{}.}
  \label{fig:overview}
\end{figure}

\begin{figure}[t]
  \centering
  \includegraphics[width=\linewidth]{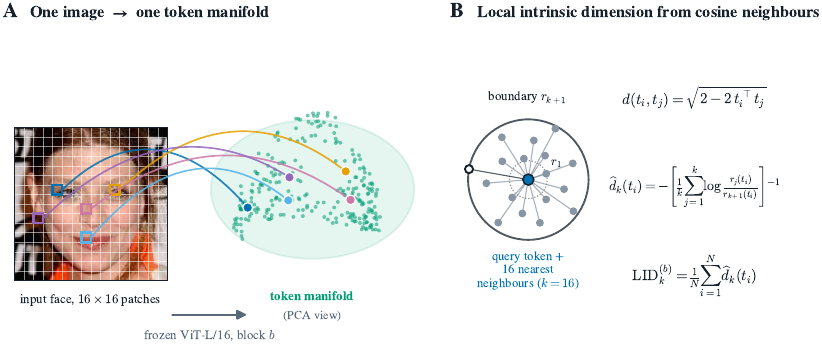}
  \caption{\textbf{The GLID signal.} \textbf{(A)}~A frozen ViT maps the $16\times16$ patches of one face to $N{=}256$ tokens; within a single image these tokens form a manifold. The cloud is a PCA view for display only --- every distance is measured in the original 1024-D feature space. \textbf{(B)}~The Levina--Bickel estimator reads the local intrinsic dimension of that manifold from each token's $k$ cosine neighbors ($r_{k+1}$ sets the local boundary; neighbor layout schematic); averaging over tokens gives one deterministic, training-free scalar per depth. No labels, no training, no reference set.}
  \label{fig:concept}
\end{figure}

\section{Introduction}
\label{sec:intro}

Face forgery has outgrown its detectors. Generators moved from GANs \citep{gan2014, stylegan2019, stylegan2_2020} to diffusion models \citep{ddpm2020, adm2021, ldm2022}, and a detector trained on classic face swaps meets diffusion portraits, one-shot reenactors, and GAN samples it has never seen. Fine-tuning a strong image backbone such as CLIP \citep{clip2021} on FaceForensics++ \citep{ffpp2019} remains the most reliable recipe \citep{effort2025}, with frozen-backbone probes close behind \citep{univfd2023}, but the recipe has a structural weakness. The model learns the artifact families present in training and stays near chance on the rest. We call these unseen families the detector's \emph{blind axes}.

This paper asks a simple question. Does a frozen foundation model already carry a signal that covers a blind axis, without any forgery labels? We answer yes, and we locate the signal in manifold geometry. The patch tokens of a single image are a sample from a low-dimensional structure inside the feature space. Generated faces perturb the \emph{local intrinsic dimension} (LID) of that structure. The perturbation is invisible to a linear head but measurable with a classical estimator \citep{levina2004}, it requires no training, and it lives at depths of the network that forensic fine-tuning barely reshapes \citep{gsd2026}.

We build this observation into \textbf{GLID} (\underline{G}ated \underline{L}ocal \underline{I}ntrinsic \underline{D}imension). GLID computes a 12-dimensional LID profile per image, from four depths and three neighborhood sizes of a frozen DINOv3 \citep{dinov3_2025}. A confidence gate injects the profile into a fine-tuned detector only where the detector is uncertain. A single scalar controls the injection strength, and we select it with an in-distribution rule that never touches evaluation data. Figure~\ref{fig:overview} summarizes the pipeline; Figure~\ref{fig:concept} details the signal.

Two empirical laws make the design principled rather than heuristic.

\textbf{Law 1: family--depth signature.} GAN-generated faces disturb the token manifold at the deepest layers. Diffusion-generated faces disturb it mid-network, and the disturbance collapses at depth. The split holds on 4 of 4 backbones, under 3 different dimension estimators, on held-out replication samples, and beyond faces (ProGAN on object imagery). No single ``best layer'' exists; the family decides the depth (Section~\ref{sec:depth}).

\textbf{Law 2: absorption boundary.} Fine-tuning absorbs an auxiliary signal's gain exactly on the axes its training distribution covers. We prove the causal version by intervention: adding 1\% (\DoseOneImages{} images) of generation-family data to training collapses the LID gain from \DoseZeroGain{} to \DoseOneGain{}, while the untouched axes keep their gains (Section~\ref{sec:absorption}). The law tells a practitioner when to use data and when to use geometry: coverage is the cheap fix for known families, GLID is the fix for unknown ones.

Our contributions:
\begin{itemize}
  \item \textbf{Method.} The first detector to use intra-image patch-manifold LID as a forgery signal, with a gated, in-distribution-calibrated injection (Section~\ref{sec:method}).
  \item \textbf{Two laws.} A family--depth signature of generators, robust across backbones, estimators, and domains; and an interventionally verified absorption boundary for auxiliary signals.
  \item \textbf{Evidence.} A 16-axis frozen benchmark with identical crops, training data, and schedules for every method; bootstrap and paired-bootstrap statistics; \NumCitations{} verified references; preregistered analysis gates reported including the one that failed.
  \item \textbf{Honesty.} GLID leads overall (\GlidAllSixteen{} vs.\ \EffortAllSixteen{} for the best rival) and never loses significantly to it on any axis (\WinEff{} wins, \LoseEff{} losses), but SBI stays stronger on the T2I axis and in the wild. We report every such case.
\end{itemize}

\section{Related work}
\label{sec:related}

\textbf{Learned forgery detectors.} The field moved from CNNs \citep{mesonet2018, ffpp2019} through frequency and blending cues \citep{f3net2020, spsl2021, facexray2020, patchforensics2020} to foundation-model probes and fine-tunes \citep{univfd2023, rine2024, effort2025}, with self-blending augmentation \citep{sbi2022} and temporal models \citep{lipforensics2021, ftcn2021, realforensics2022} as strong specializations. Up-sampling traces \citep{npr2024} and reconstruction errors \citep{dire2023, sedid2023, lare2024, aeroblade2024} target generated imagery. We retrain the strongest of these lines under one protocol.

\textbf{Training-free detection.} RIGID \citep{rigid2024} scores perturbation sensitivity of frozen embeddings; MINDER \citep{minder2024} tunes the perturbation per domain. Score-manifold curvature offers another geometric route \citep{manifoldbias2025}. DCA \citep{dca2026} measures cross-region dimension coactivation on frozen DINOv3 and is the setting closest to ours; its statistic is channel coactivation, not manifold dimension, and we show the two cover different blind axes. SPLIT \citep{split2026} and token-selection methods \citep{tokenrank2025} confirm that frozen patch tokens carry forensic signal. A recent audit questions the robustness of this family \citep{fragility2026}; we answer with corruption sweeps.

\textbf{Intrinsic dimension.} LID estimation goes back to maximum-likelihood \citep{levina2004} and TwoNN \citep{twonn2017}, with tight-locality guarantees for small samples \citep{amsaleg2022}. Depth profiles of representation ID are known \citep{ansuini2019, pope2021}. LID has served adversarial-example detection across samples \citep{ma2018lid, multilid_adv2022} and text detection via persistent homology \citep{tulchinskii2023}. multiLID \citep{multilid2023} applied inter-sample, supervised LID to generated images and was later retracted by its authors; its neighborhoods live across a reference batch, whereas ours live inside one image, unsupervised. A regularizer once aligned patch LID during inpainting \citep{plid2019}. To our knowledge no prior work scores forgeries by the intrinsic dimension of a single image's token manifold.

\textbf{Layer choice.} MoLD \citep{mold2025} observes with supervised probes that the best CLIP layer varies by dataset, and \citet{intermediate2026} searches for one optimal intermediate layer, explaining mid-layer strength with an ID hunchback. Related fine-tuning studies treat deep semantic layers as cross-domain stabilizers \citep{semanti2025}. Our family--depth signature says the premise of a single best layer is wrong: the generator family sets the depth, so a detector should read several (Section~\ref{sec:depth}).

\section{Method}
\label{sec:method}

\subsection{Intra-image patch-manifold LID}
\label{sec:lid}

Let a frozen ViT-L/16 \citep{vit2021, dinov3_2025} map a face crop (resized to the backbone's 256-px input) to $N{=}256$ patch tokens $T^{(b)}=\{t_i\}_{i=1}^{N}$ at block $b$, each $\ell_2$-normalized. We treat $T^{(b)}$ as a sample from a manifold and estimate its local intrinsic dimension with the maximum-likelihood estimator \citep{levina2004} on cosine geometry, $d(t_i,t_j)=\sqrt{2-2\,t_i^{\top}t_j}$:
\begin{equation}
  \widehat{d}_k(t_i) \;=\; -\Bigg(\frac{1}{k}\sum_{j=1}^{k}\log\frac{r_j(t_i)}{r_{k+1}(t_i)}\Bigg)^{-1},
  \qquad
  \mathrm{LID}_k^{(b)} \;=\; \frac{1}{N}\sum_{i=1}^{N}\widehat{d}_k(t_i),
  \label{eq:lid}
\end{equation}
where $r_j(t_i)$ is the distance to the $j$-th neighbor within the same image and $r_{k+1}$ defines the local boundary, matching our implementation. With 256 tokens the estimator sits in the tight-locality regime analyzed by \citet{amsaleg2022}; Section~\ref{sec:ablation} shows our conclusions are estimator-independent. The signal is a deterministic function of the input: no labels, no training, no reference set. This is the core difference from inter-sample LID pipelines \citep{ma2018lid, multilid2023}, which need a batch of other images and a supervised head.

We compute $\mathrm{LID}_k^{(b)}$ for blocks $b\in\{6,12,18,23\}$ and $k\in\{8,16,32\}$, giving a 12-dimensional profile $\ell\in\mathbb{R}^{12}$ per image at \LatLid{} ms on one GPU.

\subsection{Why several depths: the family--depth signature}
\label{sec:method-depth}

\begin{figure}[t]
  \centering
  \includegraphics[width=\linewidth]{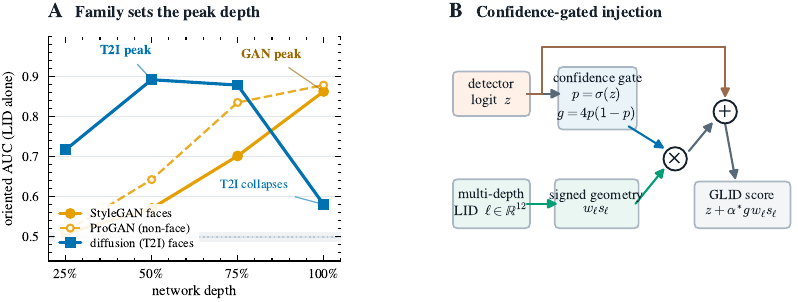}
  \caption{\textbf{Why several depths, and how the signal enters.} \textbf{(A)}~Measured LID-alone AUC across depth: diffusion (T2I) peaks mid-network and collapses at the last block, while GANs peak at the last block --- StyleGAN faces and ProGAN non-face imagery share the same pattern (\S\ref{sec:depth}). No single layer covers both families, so GLID measures a multi-depth profile. \textbf{(B)}~The profile enters a fine-tuned detector through the confidence gate $g=4p(1-p)$: on covered axes $g{\approx}0$ and the detector is preserved; on blind axes the gate opens and the signed geometry $w_\ell s_\ell$ corrects the score. The base logit always bypasses to the sum.}
  \label{fig:design}
\end{figure}

A single LID scalar would be the natural choice, and it would be wrong. Generated faces do not raise or lower dimension uniformly (Figure~\ref{fig:design}A). GAN samples disturb the manifold at the deepest block, where the profile on StyleGAN faces reaches \DinoStyleDeep{} AUC alone. Diffusion samples disturb it mid-network (\DinoTtwoiMid{} at 50\% depth) and the disturbance collapses at depth (\DinoTtwoiDeep{} at the last block). The two families are therefore mutually invisible at each other's best depth. Section~\ref{sec:depth} establishes this signature across backbones, estimators, and domains. The 12-coordinate profile is our measurement instrument and drives every analysis in Section~\ref{sec:analysis}. The deployable score below pools only the deepest and mid blocks, one per family regime; Section~\ref{sec:ablation} shows that pooling all four blocks dilutes the direction with near-chance shallow coordinates and lowers accuracy.

\subsection{Gated injection into a fine-tuned detector}
\label{sec:gate}

The base detector is a CLIP-L fine-tune (last six blocks and head) on FaceForensics++ \citep{ffpp2019}; Section~\ref{sec:protocol} gives the recipe. For an image with detector logit $B$, let $z=(B-\mu_B)/\sigma_B$ with statistics from training frames, and $p=\sigma(z)$. Standardize the LID profile the same way, $\tilde{\ell}=(\ell-\mu_\ell)/\sigma_\ell$, and pool the deep and mid coordinates: $s_\ell = \bar{\tilde{\ell}}_{23} + \bar{\tilde{\ell}}_{12}$, $w_\ell = \tanh\!\big((|\bar{\tilde{\ell}}_{23}|+|\bar{\tilde{\ell}}_{12}|)/2\big)$, where $\bar{\tilde{\ell}}_{b}$ averages the three $k$ values of block $b$. The score is
\begin{equation}
  \mathrm{GLID}(x) \;=\; z \;+\; \alpha\, \underbrace{4p(1-p)}_{\text{confidence gate } g}\; w_\ell\, s_\ell .
  \label{eq:glid}
\end{equation}
The gate $g$ opens only where the detector is unsure (Figure~\ref{fig:design}B). On covered axes the detector is confident and GLID reduces to the detector; on blind axes the geometry speaks. The magnitude weight $w_\ell$ keeps small-anomaly images quiet, and the signed pooling $s_\ell$ carries the family direction.

\subsection{Calibrating \texorpdfstring{$\alpha$}{alpha} without touching evaluation data}
\label{sec:alpha}

The remaining scalar $\alpha$ trades blind-axis gain against interference. Choosing it on evaluation axes would be tuning on test data. We instead use the only distribution a deployer owns: training frames. As $\alpha$ grows, in-distribution AUC decays monotonically (from \IndistAtZero{} at $\alpha{=}0$). We take the largest $\alpha$ whose in-distribution drop is at most $0.002$, giving $\alpha^{*}{=}\AlphaStar{}$ (in-distribution \IndistAtStar{}). Everything downstream uses this single value. Section~\ref{sec:ablation} shows the full trade-off curve; the uncalibrated choice $\alpha{=}1$ overshoots the knee and pays a reenactment cost, which explains why naive fusion attempts fail.

\section{Benchmark and protocol}
\label{sec:protocol}

\textbf{Training data.} FaceForensics++ c23 \citep{ffpp2019}, the standard cross-dataset training corpus: four manipulations plus real video. We partition the 1{,}000 identities 720/140/140 and keep a forged video only if both source and target identities fall inside one split, so no identity leaks across splits. Four frames per video give 11{,}296 training frames. Every face in the study is detected and cropped the same way (YuNet \citep{yunet2023}, margin 0.35, 224 px), because crop inconsistencies alone can move cross-dataset AUC by several points.

\textbf{Sixteen evaluation axes.} Table~\ref{tab:axes} lists the axes in five groups. \emph{Generation}: StyleGAN \citep{stylegan2019} and a mixed text-to-image pool \citep{ldm2022, sdxl2023}, against CelebA reals \citep{celeba2015}. \emph{Swap}: Celeb-DF v2 \citep{celebdf2020} with manually verified label semantics, FF++ FaceSwap, and four modern swaps from DF40 \citep{df402024}: InSwapper, SimSwap \citep{simswap2021}, FaceSwap, and BlendFace \citep{blendface2023}. \emph{Reenactment}: FF++ NeuralTextures and five DF40 reenactors \citep{fomm2019, sadtalker2023, wav2lip2020, tpsm2022}. \emph{Hard}: DFDC faces \citep{dfdc2020}. \emph{In the wild}: WildDeepfake \citep{wilddeepfake2020}. DF40 axes use only videos whose appearance identity lies outside our training split. Up to 500 images per class per axis.

\begin{table}
  \centering
  \caption{The 16 evaluation axes. Only FF++ c23 is ever trained on; every other source is unseen. DF40 fake counts are small after identity filtering; confidence intervals reflect this.}
  \label{tab:axes}
  \small
  \begin{tabular}{llrrl}
\toprule
Axis & Group & Real & Fake & Source \\
\midrule
StyleGAN & GEN & 500 & 500 & 140k real--fake faces \\
T2I & GEN & 500 & 500 & text-to-image pool / CelebA \\
CelebDF-v2 & SWAP & 500 & 500 & Celeb-DF v2 (label-verified) \\
FF-FS & SWAP & 140 & 30 & FF++ c23 \\
FF-NT & REEN & 140 & 30 & FF++ c23 \\
FOMM & REEN & 140 & 46 & DF40 (FF domain) \\
HyperReenact & REEN & 140 & 46 & DF40 (FF domain) \\
SadTalker & REEN & 140 & 46 & DF40 (FF domain) \\
Wav2Lip & REEN & 140 & 46 & DF40 (FF domain) \\
TPSM & REEN & 140 & 46 & DF40 (FF domain) \\
InSwapper & SWAP & 140 & 39 & DF40 (FF domain) \\
SimSwap & SWAP & 140 & 46 & DF40 (FF domain) \\
FaceSwap-DF40 & SWAP & 140 & 46 & DF40 (FF domain) \\
BlendFace & SWAP & 140 & 46 & DF40 (FF domain) \\
DFDC & HARD & 500 & 500 & DFDC faces \\
WildDeepfake & WILD & 500 & 500 & WildDeepfake \\
\bottomrule
\end{tabular}
\end{table}

\textbf{Baselines, retrained.} Every learned baseline uses the same backbone capacity, the same training frames, the same schedule, and the same crops: UnivFD (frozen CLIP-L probe) \citep{univfd2023}, NPR \citep{npr2024}, SBI \citep{sbi2022} on the same CLIP-L, and Effort-style orthogonal-subspace fine-tuning \citep{effort2025}. SBI, Effort, and our detector are trained with three seeds; we report per-seed spread and the mean of per-seed logits. Training-free baselines: RIGID \citep{rigid2024} with the blur perturbation recommended for faces \citep{minder2024}, our re-implementation of DCA \citep{dca2026} under identical crops (landmark-anchored regions replacing its parser), a raw SD $\epsilon$-error scorer distilled from the reconstruction family \citep{dire2023, sedid2023, lare2024}, and LID alone.

\textbf{Metrics.} Frame-level AUC per axis; 95\% bootstrap intervals; paired bootstrap for head-to-head deltas; group means and the 16-axis mean. We use raw score orientation everywhere and mark direction failures rather than silently flipping them.

\section{Results}
\label{sec:results}

\begin{table}
  \centering
  \caption{Group-mean AUC on the 16-axis benchmark. GLID uses one deployable score (Eq.~\ref{eq:glid}, $\alpha^{*}{=}\AlphaStar$). Bold marks the column best. The detector row is the 3-seed ensemble of logits that GLID builds on; the per-seed spread across 8 seeds (\BtwoSeedMean$\pm$\BtwoSeedStd{} overall) is a different estimator and is reported in the text. Full per-axis values with intervals are in Table~\ref{tab:full}.}
  \label{tab:main}
  \small
  \begin{tabular}{lcccccc}
\toprule
Method & Gen. & Swap & Reenact & DFDC & Wild & All 16 \\
\midrule
\rowcolor{gray!12} GLID (ours) & \textbf{0.816} & 0.910 & 0.720 & 0.636 & 0.830 & \textbf{0.805} \\
\midrule
CLIP-L fine-tune (3-seed ens.) & 0.731 & \textbf{0.915} & 0.725 & \textbf{0.647} & 0.825 & 0.798 \\
SBI (3-seed) & 0.742 & 0.820 & 0.663 & 0.541 & \textbf{0.870} & 0.737 \\
Effort (3-seed) & 0.726 & 0.902 & 0.716 & 0.645 & 0.833 & 0.790 \\
UnivFD & 0.606 & 0.846 & 0.679 & 0.538 & 0.717 & 0.726 \\
NPR & 0.495 & 0.564 & 0.601 & 0.519 & 0.687 & 0.575 \\
RIGID (training-free) & 0.421 & 0.539 & \textbf{0.729} & 0.552 & 0.600 & 0.600 \\
DCA (training-free) & 0.595 & 0.787 & 0.675 & 0.578 & 0.655 & 0.700 \\
LID alone (training-free) & 0.791 & 0.551 & 0.474 & 0.525 & 0.513 & 0.548 \\
\bottomrule
\end{tabular}
\end{table}

\begin{figure}
  \centering
  \includegraphics[width=\linewidth]{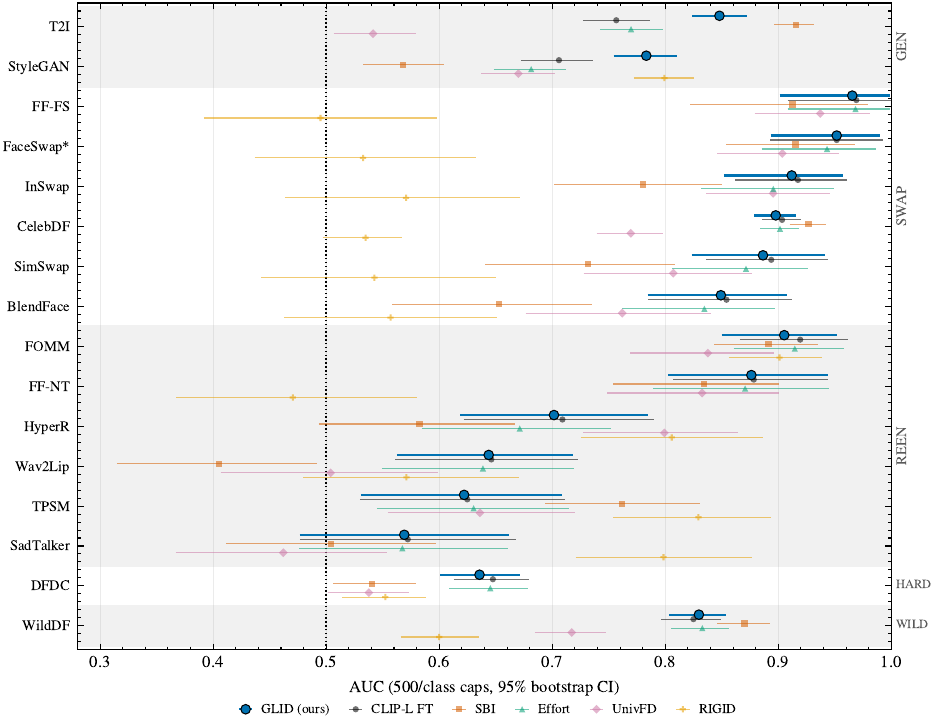}
  \caption{Per-axis AUC with 95\% bootstrap intervals for the six strongest methods. GLID (blue circles) repairs the generation axes without giving up the swap axes. DFDC stays hard for every method; WildDeepfake belongs to SBI.}
  \label{fig:bench}
\end{figure}

\textbf{Main comparison.} Table~\ref{tab:main} and Figure~\ref{fig:bench} give the headline. GLID reaches \GlidAllSixteen{} mean AUC over 16 axes, ahead of the strongest retrained rival (Effort, \EffortAllSixteen) and of its own base detector (\BtwoAllSixteen). Per-axis paired bootstrap (Table~\ref{tab:paired}) shows GLID beats Effort significantly on \WinEff{} axes and loses on \LoseEff{}. Against SBI it wins \WinSbi{} and loses \LoseSbi{}; the losses concentrate where self-blending is the right prior. The repair is not free everywhere: relative to its own base detector, GLID gives back small margins on the swap group (\GlidSwap{} vs.\ \BtwoSwap) and on DFDC (\GlidHard{} vs.\ \BtwoHard), the price of a gate that sometimes opens off-target (gate behavior in Figure~\ref{fig:gate}). The base detector itself is a strong result. Across \BtwoNSeeds{} training seeds its per-seed 16-axis mean is \BtwoSeedMean$\pm$\BtwoSeedStd, which already tops several published mechanisms under equal data. Single-seed numbers of any method can mislead by several points, so all our comparisons are multi-seed.

\textbf{Blind-spot repair.} The generation group moves from \BtwoGen{} (detector) to \GlidGen{} (\GenGain), while reenactment moves by \ReenDelta{} only. On StyleGAN the paired delta against SBI is [\PairStyleSbiLo, \PairStyleSbiHi] and against Effort [\PairStyleEffLo, \PairStyleEffHi], both excluding zero. On the T2I axis SBI remains significantly stronger ([\PairTtwoiSbiLo, \PairTtwoiSbiHi]); GLID still beats Effort there ([\PairTtwoiEffLo, \PairTtwoiEffHi]). We claim generation-group repair, not per-axis dominance.

\begin{figure}
  \centering
  \includegraphics[width=.98\linewidth]{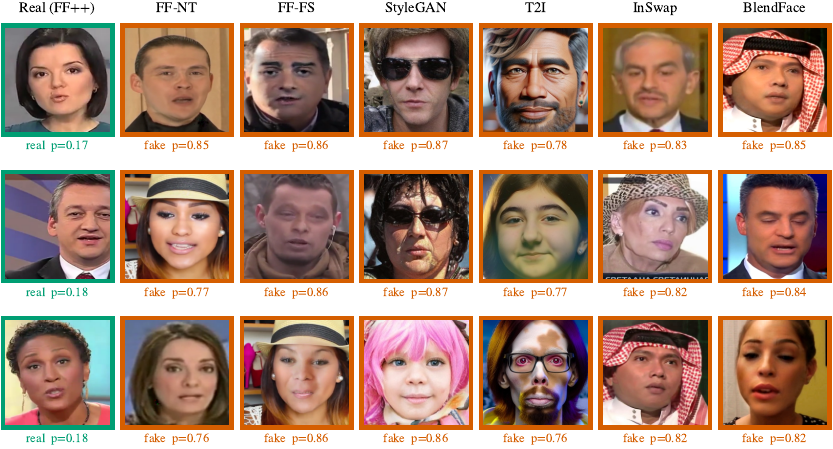}
  \caption{GLID predictions on unseen-source faces (all correct; $p$ is the predicted forgery probability). Columns: FF++ real, two FF++ manipulations, StyleGAN, text-to-image, and two DF40 swaps. Real examples come from consented FF++ actors; generated faces depict no real person.}
  \label{fig:matrix}
\end{figure}

\textbf{Qualitative behavior.} Figure~\ref{fig:matrix} shows GLID scores across sources; Appendix~\ref{app:cases} adds a larger success matrix and, importantly, the failure modes: lip-sync forgeries that edit a few mouth pixels, heavily compressed DFDC faces, and in-the-wild occlusions.

\subsection{Ablations}
\label{sec:ablation}

\begin{figure}
  \centering
  \includegraphics[width=\linewidth]{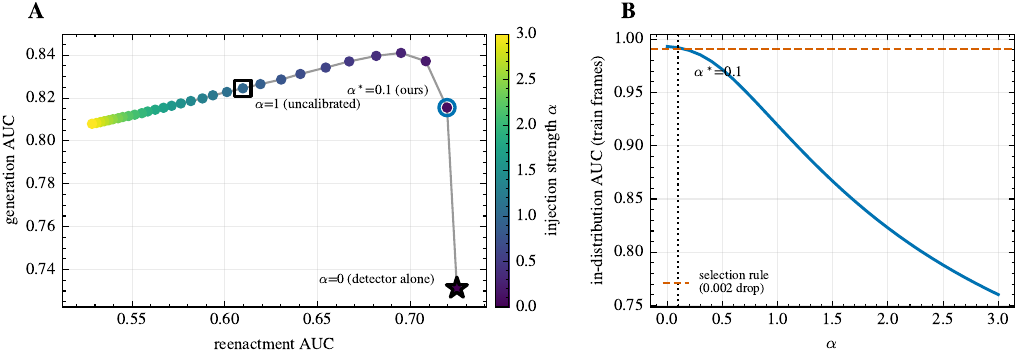}
  \caption{\textbf{(A)} Generation--reenactment trade-off as $\alpha$ sweeps; the knee sits near $\alpha{=}\KneeSoft$, and the in-distribution rule picks $\alpha^{*}{=}\AlphaStar$ without seeing any evaluation axis. $\alpha{=}1$ overshoots. \textbf{(B)} The selection rule: largest $\alpha$ with at most a $0.002$ in-distribution drop.}
  \label{fig:pareto}
\end{figure}

\textbf{Injection strength.} Figure~\ref{fig:pareto} and Table~\ref{tab:alpha} sweep $\alpha$. The trade-off has a sharp knee: at $\alpha^{*}{=}\AlphaStar$ the generation gain is nearly free, while $\alpha{=}1$ buys \AlphaOneGen{} generation at a reenactment cost down to \AlphaOneReen. All Pareto-optimal points lie in $\alpha\in[0,\ParetoMax]$. The in-distribution rule lands inside this range without evaluation access, which turns a hyperparameter into a calibration.

\begin{table}
  \centering
  \caption{Gate-strength sweep (group AUC). The in-distribution rule selects $\alpha^{*}{=}\AlphaStar$.}
  \label{tab:alpha}
  \small
  \begin{tabular}{lcccccc}
\toprule
$\alpha$ & Gen. & Swap & Reenact & DFDC & Wild & All 16 \\
\midrule
0.0 (detector) & 0.731 & 0.915 & 0.725 & 0.647 & 0.825 & 0.798 \\
0.1 ($\alpha^*$) & 0.816 & 0.910 & 0.720 & 0.636 & 0.830 & 0.805 \\
0.2 & 0.837 & 0.902 & 0.708 & 0.616 & 0.828 & 0.799 \\
0.3 & 0.841 & 0.891 & 0.695 & 0.599 & 0.819 & 0.789 \\
0.5 & 0.837 & 0.865 & 0.667 & 0.578 & 0.791 & 0.765 \\
1.0 & 0.825 & 0.818 & 0.610 & 0.557 & 0.727 & 0.719 \\
2.0 & 0.813 & 0.757 & 0.555 & 0.544 & 0.663 & 0.669 \\
\bottomrule
\end{tabular}
\end{table}

\textbf{Block pooling.} The deployable score reads one block per family regime (Eq.~\ref{eq:glid}). Pooling all four blocks instead dilutes the direction with near-chance shallow coordinates and drops overall AUC from \PoolOursAll{} to \PoolAllfourAll. Single-block variants serve one family only: block 23 alone reaches \PoolDeepGen{} on generation, block 12 alone \PoolMidGen{} (full numbers in Appendix~\ref{app:constants}).

\begin{figure}
  \centering
  \includegraphics[width=\linewidth]{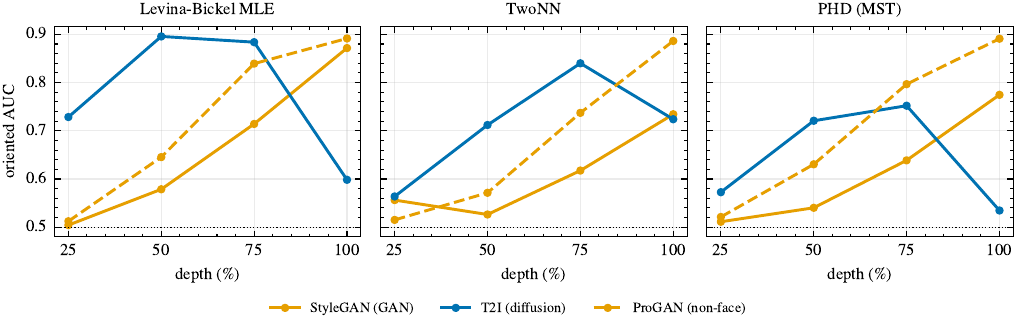}
  \caption{The family--depth signature is estimator-independent. Three intrinsic-dimension estimators on the same token clouds reproduce the split: diffusion peaks mid-depth and collapses at the last block, GAN and non-face ProGAN peak at the deepest block.}
  \label{fig:estimators}
\end{figure}

\textbf{Estimator choice.} Replacing the MLE with TwoNN \citep{twonn2017} or a persistent-homology-style MST estimator \citep{tulchinskii2023} preserves the family--depth signature on all three generation sources (Figure~\ref{fig:estimators}); the MLE is uniformly strongest, which justifies its use.

\textbf{Backbone choice.} The signature and the signal survive DINOv2 \citep{dinov2_2023}, CLIP \citep{clip2021}, and EVA-02 \citep{eva02_2023} (Appendix Figure~\ref{fig:depthheat}); self-supervised backbones carry the strongest geometry, consistent with \citet{rigid2024}.

\section{Analysis}
\label{sec:analysis}

\subsection{The absorption boundary, established by intervention}
\label{sec:absorption}

Why does a 12-number signal beat retraining on blind axes, yet add nothing elsewhere? Our answer is a law about fine-tuning, not about faces. Prior work shows fine-tuning distorts pretrained features \citep{kumar2022lpft}, that dominant features starve alternatives of gradient \citep{gradstarve2021}, and that forensic fine-tunes leave representations largely semantic \citep{gsd2026}. We state the consequence for signal fusion and test it causally:

\begin{quote}
\emph{A fine-tuned detector absorbs an auxiliary signal's gain exactly on the axes its training distribution covers; on uncovered axes the gain survives.}
\end{quote}

\begin{figure}
  \centering
  \includegraphics[width=\linewidth]{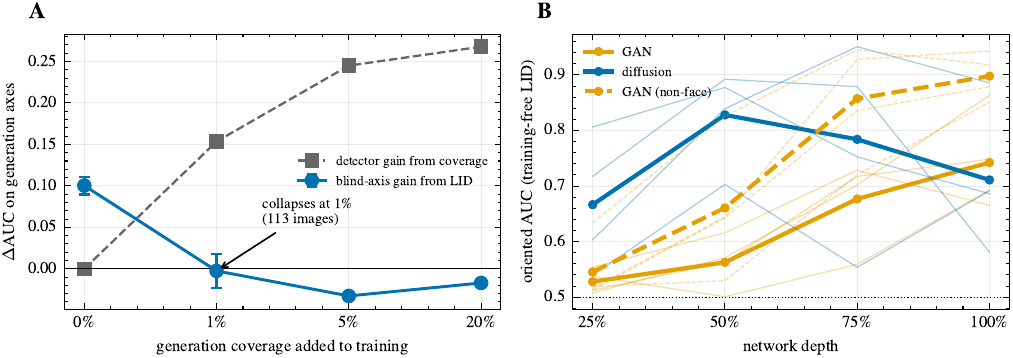}
  \caption{\textbf{(A)} The absorption boundary, interventionally. Adding generation coverage to training collapses the blind-axis LID gain at 1\% already, while the detector's own accuracy on that family rises. \textbf{(B)} The family--depth signature. Thin lines: four backbones; thick: mean. GAN families (face and non-face) peak at the deepest block; diffusion peaks mid-network and collapses at depth.}
  \label{fig:hero}
\end{figure}

The interventional test injects a controlled dose of generation-family images into the training set and retrains (three seeds per dose). Figure~\ref{fig:hero}A and Table~\ref{tab:dose} give the dose--response. At 0\% coverage the LID gain on generation axes is \DoseZeroGain$\pm$\DoseZeroStd. At 1\% coverage (\DoseOneImages{} images) the gain is \DoseOneGain: gone. The detector's own generation AUC jumps from \DoseZeroBtwoGen{} to \DoseOneBtwoGen{} and reaches \DoseTwentyBtwoGen{} at 20\%. Reenactment gains stay flat within seed noise at every dose. Absorption is threshold-like, not gradual.

The law has two practical corollaries. First, if you can name the family, buy data, not machinery: 1\% coverage outperforms any auxiliary signal we measured. Second, auxiliary mechanisms must be evaluated on axes outside training coverage, or absorption will erase them and the evaluation will call them useless. The same law explains our own negative result: a preregistered variant that fused LID with a learned global head failed its gates because the head, fitted on covered data, learned the signal away (Appendix~\ref{app:prereg}).

\subsection{The family--depth signature is a property of generators}
\label{sec:depth}

Figure~\ref{fig:hero}B overlays depth profiles for four backbones on three generation sources. The GAN profile rises monotonically to the last block (StyleGAN faces; ProGAN objects \citep{progan2018, cnndetect2020}), the diffusion profile peaks at 50--75\% and collapses at depth. Three estimators agree (Figure~\ref{fig:estimators}). Because ProGAN images contain no faces, the signature tracks the generator family rather than the face domain. This resolves the single-best-layer question raised by supervised layer studies \citep{mold2025, intermediate2026}: the optimum is family-conditional, so any single-layer detector is blind to one family by construction. It also mirrors depth profiles of representation dimensionality in classification \citep{ansuini2019, pope2021}, now with a forensic direction attached.

\subsection{Feature geometry at a glance}

\begin{figure}
  \centering
  \includegraphics[width=.5\linewidth]{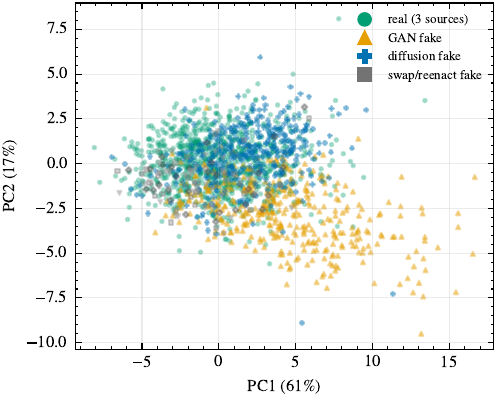}
  \caption{PCA of LID profiles (fitted on real faces). GAN and diffusion fakes exit the real cloud along distinct directions; swap and reenactment fakes remain inside, which is exactly where the fine-tuned detector is already strong.}
  \label{fig:pca}
\end{figure}

Figure~\ref{fig:pca} projects the 12-dimensional LID profiles with PCA fitted on real faces only. Generation families leave the real cloud along family-specific directions; swap and reenactment forgeries stay inside it. The picture explains both the power and the limit of the signal in one plot: geometry separates what generators synthesize, not what editors paste.

\subsection{The training-free triangle and direction stability}
\label{sec:triangle}

Training-free detectors split the blind axes among themselves: DCA is swap-oriented (swap \DcaSwap, generation \DcaGen; our re-implementation reaches \DcaCelebdf{} on Celeb-DF), LID is generation-oriented (generation \LidGen, swap \LidSwap), and RIGID is reenactment-oriented (reenactment \RigidReen). Naive score or feature fusion of DCA and LID fails to dominate both parents; the fusion head, fitted on covered data, reproduces absorption in miniature. Covering multiple blind axes without labels remains open.

\begin{figure}
  \centering
  \includegraphics[width=\linewidth]{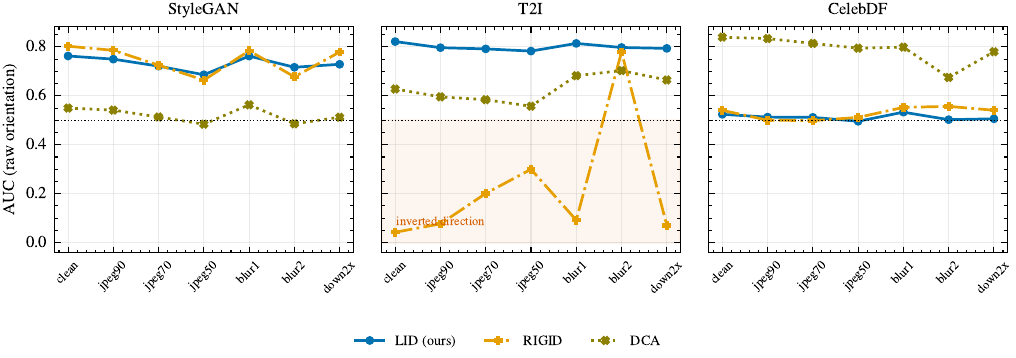}
  \caption{Corruption sweeps for three training-free signals. LID degrades smoothly and never flips direction; RIGID's direction is unstable on T2I (shaded region marks inversion); DCA is robust on its home axis (Celeb-DF).}
  \label{fig:robust}
\end{figure}

Direction stability separates the geometric signal from the sensitivity family. RIGID's raw direction inverts across axes (T2I raw AUC \RigidTtwoiRaw) and wanders under corruption, reaching \RigidTtwoiJpegfifty{} at JPEG-50 and \RigidTtwoiBlur{} under strong blur. The raw SD $\epsilon$-error inverts on T2I as well (\LareTtwoiRaw): generated images sit close to the generator's own manifold. A signed scorer whose sign flips per axis cannot be deployed without per-axis knowledge. The LID direction is family-consistent and degrades smoothly (T2I \LidTtwoiClean{} clean, \LidTtwoiJpegfifty{} at JPEG-50; Figure~\ref{fig:robust}), answering the robustness audit of \citet{fragility2026} for this signal.

\subsection{Deterministic signals stabilize seeds}
\label{sec:variance}

\begin{figure}
  \centering
  \includegraphics[width=.9\linewidth]{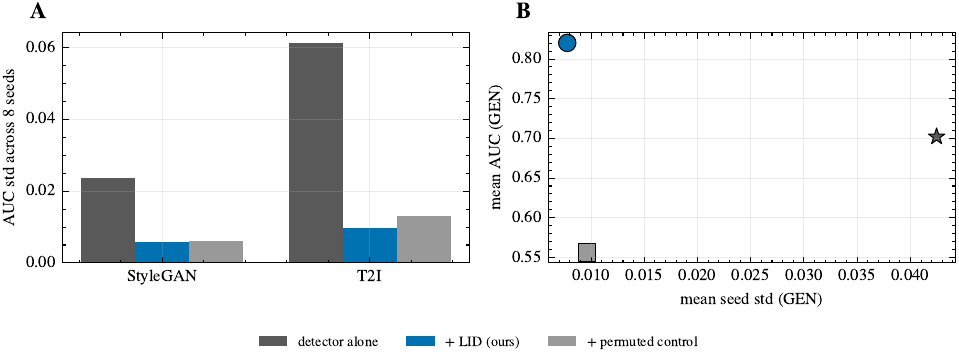}
  \caption{\textbf{(A)} Seed standard deviation on generation axes: detector, detector+LID, and a permuted deterministic control. \textbf{(B)} Mean accuracy against seed spread: only LID moves up and left.}
  \label{fig:variance}
\end{figure}

Across \BtwoNSeeds{} training seeds, the detector's generation-axis AUC varies with standard deviation \VarBtwoGen. Adding the LID injection cuts it to \VarLidGen{} (\VarRatioLid$\times$). A control that injects the same magnitudes with permuted image assignment also stabilizes (\VarCtrlGen, \VarRatioCtrl$\times$) but destroys accuracy (\VarGainCtrl{} vs.\ \VarGainLid{} for LID). Stabilization is therefore a generic property of deterministic injection; the contribution of LID is that gain and stability arrive together (Figure~\ref{fig:variance}). Seed variance of fine-tuned detectors is rarely reported \citep{mosbach2021, picard2021}; at these magnitudes it can decide leaderboard order, which is why every learned method here is multi-seed.

\subsection{Cost and calibration}
\label{sec:cost}

The LID signal adds \LatLid{} ms per image to a \LatBtwo{} ms detector (batch 32, one GPU), cheaper than RIGID (\LatRigid{} ms) and DCA (\LatDca{} ms). With temperature $T{=}\EceT$ fitted on validation frames, expected calibration error is \EceValT{} in-distribution but \EceCrossT{} averaged across the 16 axes \citep{guo2017calib}: cross-generator probabilities remain miscalibrated even when ranking is good, a caution for downstream use.

\section{Limitations}
\label{sec:limits}

Face-domain generation sources are thin: one GAN corpus and one mixed T2I pool; the cross-domain and cross-estimator checks mitigate but do not remove this. On T2I, self-blending training remains significantly stronger than our repair, and in-the-wild data favors SBI (\SbiWild). DFDC stays near \GlidHard{} for every method we measured; heavy compression suppresses both semantic and geometric evidence. The 256-token estimator is biased at small $k$ \citep{amsaleg2022}, which we absorb into direction conventions rather than correct. Our protocol is frame-level; video aggregation would lift absolute numbers. Finally, GLID repairs generation blind axes specifically; lip-sync micro-edits remain open (Appendix~\ref{app:cases}).

\section{Conclusion}

GLID turns a classical quantity, local intrinsic dimension, into a deployable repair for the blind axes of fine-tuned face-forgery detectors. The method is one equation on top of a frozen backbone. Its justification is two laws with unusual evidential support for this field: a family--depth signature that survives backbones, estimators, and domains, and an absorption boundary verified by intervention. We hope the second law changes practice beyond this paper: auxiliary signals should be measured where training coverage ends, because that is the only place they can matter.

\section*{Acknowledgments}

We thank the Phi Lab Foundation for providing the computing infrastructure and the research funding that supported this work.

\bibliographystyle{unsrtnat}
\bibliography{references}

\clearpage
\appendix
\raggedbottom  

\section{Success and failure cases}
\label{app:cases}

\begin{figure}[H]
  \centering
  \includegraphics[width=.9\linewidth]{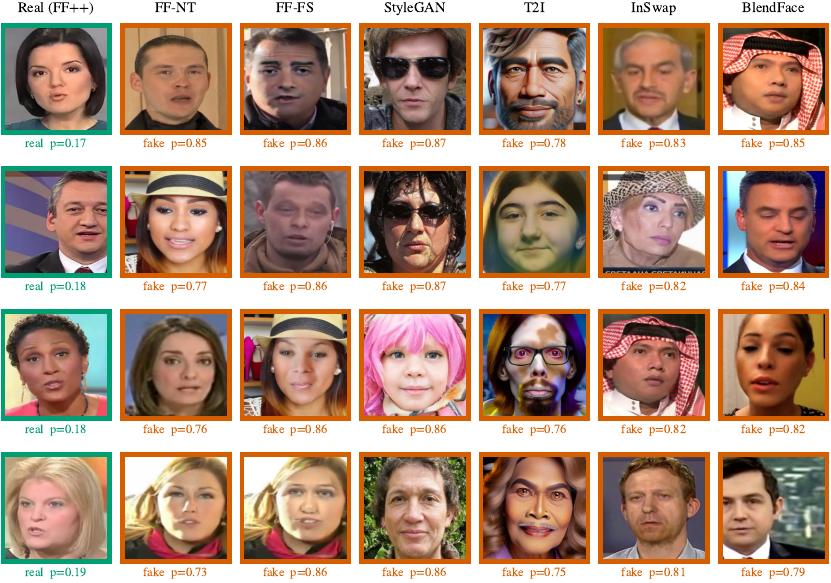}\\[2pt]
  {\footnotesize\textbf{(A)} Success cases}\\[8pt]
  \includegraphics[width=.9\linewidth]{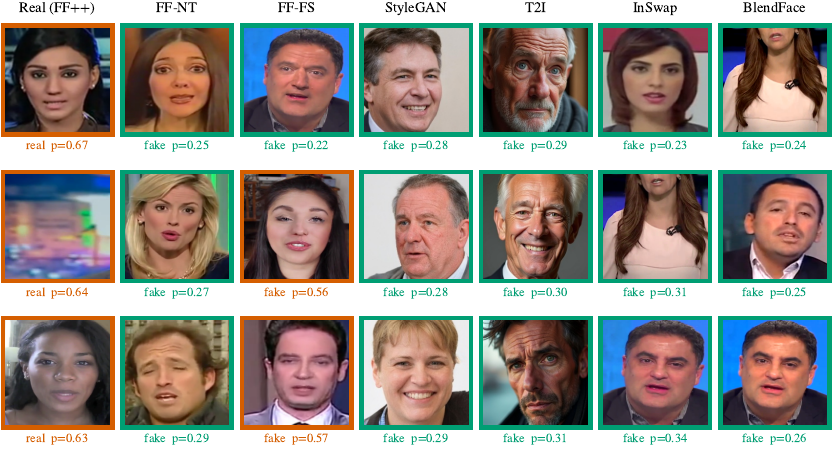}\\[2pt]
  {\footnotesize\textbf{(B)} Failure cases}
  \caption{\textbf{(A)} Extended success matrix: four examples per source, all correctly scored by GLID. \textbf{(B)} Failure cases: the most confident mistakes per source. Three patterns dominate. (1) Reenactment and lip-sync forgeries that alter a small mouth region leave both the semantic features and the token manifold nearly intact. (2) Strong compression (DFDC-like) flattens the geometric evidence. (3) Occlusions, heavy makeup, and sunglasses push real faces toward the forged score region.}
  \label{fig:successlarge}
\end{figure}

Reading the failures through the method explains them. The LID signal measures how synthesis bends an image's global token manifold, so forgeries that paste or warp a small region (Wav2Lip, some reenactors) barely move it; the gate then defers to the detector, which is also weak there. Compression removes the high-frequency support of the manifold measurement. These are the axes where neither data coverage nor geometry currently helps, and we flag them as the open end of the problem.

\section{Preregistered analysis gates}
\label{app:prereg}

We fixed decision gates before running the corresponding experiments and report all of them.

\begin{itemize}
  \item \textbf{Signal viability.} Training-free LID must exceed 0.85 AUC on at least one generation source under our crops; below 0.75 the line is abandoned. Outcome: passed.
  \item \textbf{Unified learned fusion.} A single learned head over detector and geometric features must gain on generation and reenactment simultaneously with in-distribution regression under 0.01. Outcome: failed; the head absorbed the geometric signal (Section~\ref{sec:absorption}), and we report it as a negative result rather than tuning past it. The gated form of Eq.~\ref{eq:glid} with in-distribution $\alpha$ replaced it.
  \item \textbf{Absorption intervention.} Success defined as gain collapse to at most $+0.01$ at 20\% coverage with reenactment unchanged within $0.02$ and rising covered-axis accuracy. Outcome: passed in sharpened form; collapse completes at 1\% (Table~\ref{tab:dose}), and the strict monotonicity sub-check is moot because collapse is immediate. The reenactment-shift sub-check missed its threshold by \ReenMissBy, which we report as-is.
  \item \textbf{Depth-signature breadth.} The split must hold for at least 3 of 4 backbones and all 3 estimators on every valid source. Outcome: passed (the EVA-02 ProGAN peak and the CLIP StyleGAN peak sit at 75\% depth with the last-block values close behind; all other combinations peak at the last block).
  \item \textbf{Replication.} Direction of the depth split re-verified on disjoint, freshly sampled evaluation images before any comparison was finalized. Outcome: passed.
\end{itemize}

\section{Protocol constants}
\label{app:constants}

\textbf{Crops.} YuNet detection with margin 0.35 at 224 px for every image in training and evaluation. The margin matches the training pipeline of the base detector; mixing crop styles across methods moved AUC by up to \CropDelta{} in our checks, so we forbid it.

\textbf{Splits.} Identities 720/140/140; forged videos kept only when both endpoint identities share a split. DF40 evaluation uses appearance identities outside the training split; this leaves 39--46 fake videos for the swap axes (Table~\ref{tab:axes}) and widens their intervals.

\textbf{Detector.} CLIP-L/14, last 6 blocks and a LayerNorm head trainable, AdamW, lr $10^{-5}$ backbone and $10^{-4}$ head, weight decay $0.01$, batch 32, two epochs, class-balanced sampling, horizontal flips; best-validation checkpoint per seed. SBI and Effort use the same capacity and schedule. NPR trains its reference ResNet-18. UnivFD sweeps its regularizer on a held-out tenth of training.

\textbf{LID.} Blocks $\{6,12,18,23\}$, $k\in\{8,16,32\}$, cosine distances on $\ell_2$-normalized tokens. Statistics $\mu_\ell, \sigma_\ell, \mu_B, \sigma_B$ come from training frames only.

\textbf{Gate pooling.} The deployable score pools blocks 23 and 12 (Eq.~\ref{eq:glid}). At $\alpha^{*}$ this choice gives \PoolOursGen{} generation and \PoolOursAll{} overall. Pooling all four blocks adds the near-chance shallow coordinates and drops to \PoolAllfourGen{} generation (\PoolAllfourAll{} overall). Single-block pooling serves one family only: block 23 alone gives \PoolDeepGen{} generation, block 12 alone \PoolMidGen{}. Reading one depth per family regime is both necessary and sufficient.

\textbf{DCA re-implementation.} Frozen DINOv3 block 18 at 448 px, per-dimension normalization from training frames, $K{=}20$ tokens per region, eye--mouth--nose pairs, logistic probe with $C{=}0.1$; regions anchored on detector landmarks, replacing the original RetinaFace-plus-parser stack \citep{retinaface2020, farl2022} so that every method consumes identical crops.

\section{Full tables}

\begin{table}[H]
  \centering
  \caption{Per-axis AUC with 95\% bootstrap intervals.}
  \label{tab:full}
  \scriptsize
  \setlength{\tabcolsep}{2.5pt}
  \resizebox{\textwidth}{!}{\begin{tabular}{lcccccccc}
\toprule
Axis & GLID (ours) & CLIP-L FT & SBI & Effort & UnivFD & NPR & RIGID & DCA \\
\midrule
StyleGAN & 0.783\tiny{[0.75,0.81]} & 0.706\tiny{[0.67,0.74]} & 0.568\tiny{[0.53,0.60]} & 0.681\tiny{[0.65,0.71]} & 0.670\tiny{[0.64,0.70]} & 0.534\tiny{[0.50,0.57]} & 0.799\tiny{[0.77,0.83]} & 0.565\tiny{[0.53,0.60]} \\
T2I & 0.848\tiny{[0.82,0.87]} & 0.757\tiny{[0.73,0.79]} & 0.915\tiny{[0.90,0.93]} & 0.770\tiny{[0.74,0.80]} & 0.541\tiny{[0.51,0.58]} & 0.457\tiny{[0.42,0.49]} & 0.043\tiny{[0.03,0.06]} & 0.625\tiny{[0.59,0.66]} \\
CelebDF-v2 & 0.898\tiny{[0.88,0.92]} & 0.903\tiny{[0.89,0.92]} & 0.927\tiny{[0.91,0.94]} & 0.902\tiny{[0.88,0.92]} & 0.770\tiny{[0.74,0.80]} & 0.699\tiny{[0.67,0.73]} & 0.535\tiny{[0.50,0.57]} & 0.847\tiny{[0.82,0.87]} \\
FF-FS & 0.965\tiny{[0.90,1.00]} & 0.969\tiny{[0.91,1.00]} & 0.913\tiny{[0.82,0.98]} & 0.968\tiny{[0.91,1.00]} & 0.937\tiny{[0.88,0.98]} & 0.515\tiny{[0.41,0.61]} & 0.495\tiny{[0.39,0.60]} & 0.837\tiny{[0.74,0.92]} \\
FF-NT & 0.876\tiny{[0.80,0.94]} & 0.878\tiny{[0.81,0.94]} & 0.834\tiny{[0.75,0.90]} & 0.871\tiny{[0.79,0.95]} & 0.833\tiny{[0.75,0.90]} & 0.499\tiny{[0.39,0.61]} & 0.470\tiny{[0.37,0.58]} & 0.774\tiny{[0.68,0.86]} \\
FOMM & 0.905\tiny{[0.85,0.95]} & 0.919\tiny{[0.87,0.96]} & 0.891\tiny{[0.84,0.94]} & 0.915\tiny{[0.86,0.96]} & 0.838\tiny{[0.77,0.90]} & 0.705\tiny{[0.62,0.78]} & 0.901\tiny{[0.86,0.94]} & 0.736\tiny{[0.65,0.82]} \\
HyperReenact & 0.702\tiny{[0.62,0.78]} & 0.709\tiny{[0.62,0.79]} & 0.582\tiny{[0.49,0.67]} & 0.671\tiny{[0.58,0.75]} & 0.799\tiny{[0.73,0.86]} & 0.571\tiny{[0.48,0.66]} & 0.806\tiny{[0.73,0.89]} & 0.703\tiny{[0.61,0.78]} \\
SadTalker & 0.569\tiny{[0.48,0.66]} & 0.572\tiny{[0.48,0.67]} & 0.505\tiny{[0.41,0.60]} & 0.567\tiny{[0.48,0.66]} & 0.462\tiny{[0.37,0.55]} & 0.607\tiny{[0.51,0.70]} & 0.798\tiny{[0.72,0.88]} & 0.549\tiny{[0.46,0.64]} \\
Wav2Lip & 0.644\tiny{[0.56,0.72]} & 0.646\tiny{[0.56,0.72]} & 0.405\tiny{[0.31,0.49]} & 0.639\tiny{[0.55,0.72]} & 0.504\tiny{[0.41,0.60]} & 0.522\tiny{[0.42,0.61]} & 0.571\tiny{[0.48,0.67]} & 0.686\tiny{[0.60,0.76]} \\
TPSM & 0.622\tiny{[0.53,0.71]} & 0.625\tiny{[0.53,0.71]} & 0.761\tiny{[0.69,0.83]} & 0.630\tiny{[0.54,0.71]} & 0.636\tiny{[0.55,0.72]} & 0.705\tiny{[0.62,0.78]} & 0.829\tiny{[0.75,0.89]} & 0.600\tiny{[0.50,0.70]} \\
InSwapper & 0.912\tiny{[0.85,0.96]} & 0.917\tiny{[0.86,0.96]} & 0.781\tiny{[0.70,0.85]} & 0.896\tiny{[0.83,0.95]} & 0.895\tiny{[0.84,0.95]} & 0.588\tiny{[0.48,0.68]} & 0.571\tiny{[0.46,0.67]} & 0.726\tiny{[0.64,0.82]} \\
SimSwap & 0.886\tiny{[0.82,0.94]} & 0.894\tiny{[0.84,0.94]} & 0.731\tiny{[0.64,0.81]} & 0.871\tiny{[0.81,0.93]} & 0.807\tiny{[0.73,0.88]} & 0.473\tiny{[0.38,0.57]} & 0.543\tiny{[0.44,0.65]} & 0.780\tiny{[0.70,0.86]} \\
FaceSwap-DF40 & 0.952\tiny{[0.89,0.99]} & 0.952\tiny{[0.89,0.99]} & 0.915\tiny{[0.85,0.97]} & 0.943\tiny{[0.89,0.99]} & 0.904\tiny{[0.85,0.95]} & 0.481\tiny{[0.38,0.57]} & 0.533\tiny{[0.44,0.63]} & 0.812\tiny{[0.73,0.88]} \\
BlendFace & 0.849\tiny{[0.78,0.91]} & 0.854\tiny{[0.78,0.91]} & 0.653\tiny{[0.56,0.73]} & 0.835\tiny{[0.76,0.90]} & 0.762\tiny{[0.68,0.84]} & 0.629\tiny{[0.54,0.72]} & 0.557\tiny{[0.46,0.65]} & 0.718\tiny{[0.63,0.80]} \\
DFDC & 0.636\tiny{[0.60,0.67]} & 0.647\tiny{[0.61,0.68]} & 0.541\tiny{[0.51,0.58]} & 0.645\tiny{[0.61,0.68]} & 0.538\tiny{[0.50,0.57]} & 0.519\tiny{[0.49,0.56]} & 0.552\tiny{[0.51,0.59]} & 0.578\tiny{[0.54,0.61]} \\
WildDeepfake & 0.830\tiny{[0.80,0.85]} & 0.825\tiny{[0.80,0.85]} & 0.870\tiny{[0.85,0.89]} & 0.833\tiny{[0.81,0.86]} & 0.717\tiny{[0.68,0.75]} & 0.687\tiny{[0.65,0.72]} & 0.600\tiny{[0.57,0.64]} & 0.655\tiny{[0.62,0.69]} \\
\bottomrule
\end{tabular}}
\end{table}

\begin{table}[H]
  \centering
  \caption{Paired bootstrap deltas of GLID against the two strongest retrained baselines. Asterisks mark intervals excluding zero.}
  \label{tab:paired}
  \small
  \begin{tabular}{lcc}
\toprule
Axis & vs.\ SBI & vs.\ Effort \\
\midrule
StyleGAN & [+0.188,+0.244]$^{*}$ & [+0.086,+0.120]$^{*}$ \\
T2I & [-0.087,-0.049]$^{*}$ & [+0.063,+0.095]$^{*}$ \\
CelebDF-v2 & [-0.047,-0.010]$^{*}$ & [-0.012,+0.004] \\
FF-FS & [+0.010,+0.115]$^{*}$ & [-0.009,+0.001] \\
FF-NT & [-0.039,+0.130] & [-0.016,+0.026] \\
FOMM & [-0.035,+0.063] & [-0.026,+0.007] \\
HyperReenact & [+0.029,+0.212]$^{*}$ & [+0.004,+0.059]$^{*}$ \\
SadTalker & [-0.023,+0.166] & [-0.041,+0.041] \\
Wav2Lip & [+0.146,+0.334]$^{*}$ & [-0.039,+0.044] \\
TPSM & [-0.226,-0.059]$^{*}$ & [-0.042,+0.024] \\
InSwapper & [+0.063,+0.204]$^{*}$ & [-0.013,+0.049] \\
SimSwap & [+0.082,+0.227]$^{*}$ & [-0.010,+0.041] \\
FaceSwap-DF40 & [-0.008,+0.086] & [-0.001,+0.023] \\
BlendFace & [+0.119,+0.280]$^{*}$ & [-0.011,+0.040] \\
DFDC & [+0.060,+0.128]$^{*}$ & [-0.027,+0.007] \\
WildDeepfake & [-0.067,-0.014]$^{*}$ & [-0.010,+0.004] \\
\bottomrule
\end{tabular}
\end{table}

\begin{table}[H]
  \centering
  \caption{Coverage dose--response (three seeds per dose).}
  \label{tab:dose}
  \small
  \begin{tabular}{lcccc}
\toprule
Coverage & Gen.\ gain from LID & Reenact gain & Detector Gen.\ AUC & All-16 AUC \\
\midrule
0\% & +0.100$\pm$0.011 & -0.123$\pm$0.012 & 0.720 & 0.792 \\
1\% & -0.003$\pm$0.020 & -0.103$\pm$0.014 & 0.873 & 0.799 \\
5\% & -0.033$\pm$0.001 & -0.107$\pm$0.006 & 0.965 & 0.806 \\
20\% & -0.017$\pm$0.002 & -0.101$\pm$0.007 & 0.988 & 0.822 \\
\bottomrule
\end{tabular}
\end{table}

\begin{table}[H]
  \centering
  \caption{Inference cost per component.}
  \label{tab:latency}
  \small
  \begin{tabular}{lcc}
\toprule
Component & ms/image & Notes \\
\midrule
Detector (CLIP-L) & 3.6 & batch 32, H100 \\
LID signal & 4.1 & 12-d, frozen ViT-L \\
RIGID & 8.0 & two forward passes \\
DCA & 18.7 & 448 px, region pairs \\
SD $\epsilon$-error & 14.2 & two timesteps \\
\bottomrule
\end{tabular}
\end{table}

\section{Additional figures}

\begin{figure}[H]
  \centering
  \includegraphics[width=.9\linewidth]{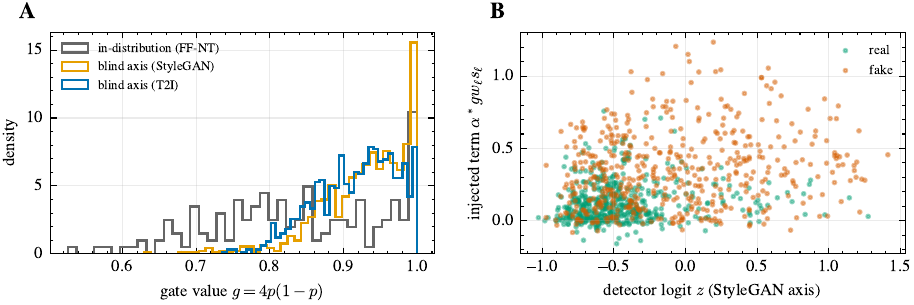}
  \caption{Gate behavior. \textbf{(A)} The confidence gate opens far more often on blind axes than in-distribution. \textbf{(B)} Injected term against detector logit on the StyleGAN axis: the injection concentrates where the detector is uncertain and pushes fakes upward.}
  \label{fig:gate}
\end{figure}

\begin{figure}[H]
  \centering
  \includegraphics[width=\linewidth]{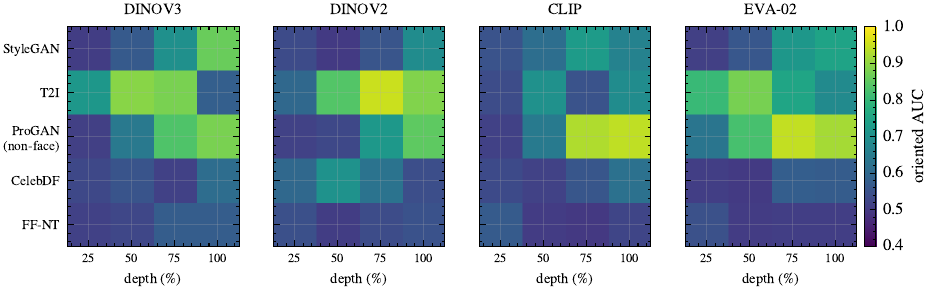}
  \caption{Depth--axis AUC heatmaps for four backbones. The family--depth signature (GAN deep, diffusion mid) repeats in every panel; swap and reenactment stay near chance at every depth, as expected for a generation-oriented signal.}
  \label{fig:depthheat}
\end{figure}

\begin{figure}[H]
  \centering
  \includegraphics[width=\linewidth]{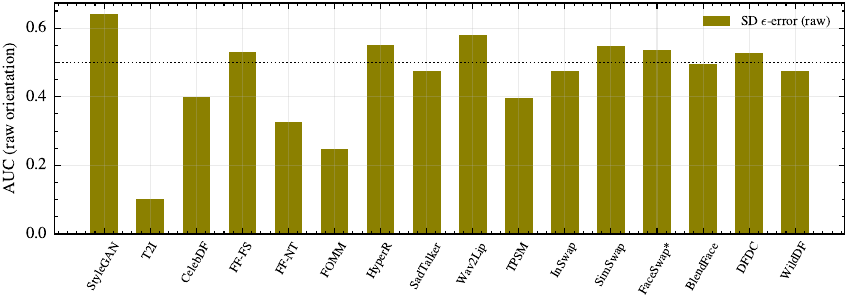}
  \caption{Raw SD $\epsilon$-error as a single scorer across the 16 axes. The direction inverts on T2I (values far below 0.5): images from the generator family of the scoring model reconstruct too well. Direction instability, not weak magnitude, is what disqualifies reconstruction scores as deployable single detectors.}
  \label{fig:lare}
\end{figure}

\section{Reproducibility}
\label{app:repro}

All experiments run in one container image with pinned library versions; training and evaluation scripts, preregistered gates, per-image score dumps, and the exact figure-generation code ship with the paper. Every number in the text and every figure is generated from one machine-readable extract of the result files; no value is typed by hand. Face examples in Figures~\ref{fig:matrix} and \ref{fig:successlarge} are shown under the FaceForensics++ terms of use (consented actors) and the CelebA research license; generated faces depict no real person. Total compute for the study, including all baselines, seeds, and interventions, is under six GPU-days on H100-class hardware.

\section{Broader impact}
\label{app:impact}

Forgery detection is a defensive technology, and better blind-axis coverage directly serves platforms, journalists, and forensic analysts who face generators they have never seen. Two risks deserve statement. First, publishing the LID signal tells adversaries what to attack; a generator regularized toward real-face token dimensionality could evade it, which is one more reason to deploy geometric signals alongside learned detectors rather than instead of them. Second, detector scores are not calibrated across generator families (Section~\ref{sec:cost}), so downstream users should not read the probabilities as evidence strength without recalibration. All face data in this study comes from public research datasets under their licenses, and no new personal data was collected.


\end{document}